\documentclass[runningheads]{llncs}

\usepackage[T1]{fontenc}
\usepackage{graphicx}
\usepackage{booktabs}
\usepackage{multirow}
\usepackage{amsmath,amssymb,amsfonts}
\usepackage{bm}
\usepackage{siunitx}
\usepackage{subcaption}
\usepackage{xcolor}
\usepackage{hyperref}
\usepackage{enumitem}
\usepackage{url}
\usepackage{tikz}
\usetikzlibrary{shapes,arrows,positioning}
\usepackage{float}

\hypersetup{colorlinks=true,linkcolor=blue,citecolor=blue,urlcolor=blue}
\begin{document}

\title{Quantum Reservoir Computing with Physics-Informed Correction for Reduced-Order PDE Forecasting}
\titlerunning{QRC with physics-informed corrector for PDE}

\author{
Krishna Bhatia\orcidID{0009-0008-2436-9853}\inst{1},
Harsh\inst{2} \and Shalini Devendrababu\inst{1}
}

\authorrunning{Krishna, Harsh and Shalini}

\institute{
QuantumAI Lab, Fractal Analytics, Mumbai, India \and
Department of Physics, Indian Institute of Technology, Jodhpur, India
}

\maketitle

\begin{abstract}
We study a hybrid proposal--correction architecture for reduced-order PDE forecasting in which a pure-state quantum reservoir computer (QRC) predicts latent coefficient dynamics and a PINN-based physics-informed corrector (PIC) refines local rollout windows. The method is evaluated on Burgers and Kuramoto--Sivashinsky (KS), with KS as the primary chaotic benchmark. On KS, QRC+PIC consistently improves over QRC alone in RMSE, NRMSE, and PDE residual, while Burgers highlights a regime in which simple baselines remain strong. These results suggest that QRC proposals with local physics-informed correction are a viable benchmark-dependent reduced-order forecasting strategy.

\end{abstract}

\section{Introduction}
Reduced-order forecasting of nonlinear and chaotic PDEs sits at the intersection of dynamical systems, numerical simulation, and machine learning. Canonical examples such as the viscous Burgers equation and the Kuramoto--Sivashinsky (KS) equation expose complementary difficulties: Burgers combines nonlinear advection and diffusion with steep gradients, whereas KS adds spatiotemporal chaos and strong sensitivity to rollout error \cite{raissi2019pinns,pathak2018ks,pinnacle2024}. This combination makes them useful testbeds for hybrid scientific machine learning methods.

Reservoir computing (RC) is attractive in this setting because it shifts most optimization effort to a linear readout while retaining nonlinear state evolution in the reservoir \cite{jaeger2001echo,pathak2018ks}. Physics-informed neural networks (PINNs), by contrast, incorporate governing equations as soft constraints and can reduce nonphysical drift, but their optimization can become fragile for convection-dominated or chaotic systems \cite{raissi2019pinns,karniadakis2021piml,krishnapriyan2021failure}. Recent physics-informed RC variants, including PI-ESN and API-ESN, show that modest physical regularization can extend predictability in chaotic settings without abandoning efficient sequence models \cite{doan2020piesn,racca2021apiesn}. In parallel, the QRC literature has argued that compact quantum reservoirs can be useful nonlinear feature maps for forecasting chaotic dynamics and reduced-order fluid systems \cite{fujii2017qrc,ahmed2024extreme,ahmed2025robust,pfeffer2022thermal,pfeffer2023rbcrom}.

This paper investigates a conservative hybridization of these threads. Rather than replacing the entire forecasting stack with a quantum or physics-informed model, we study a two-stage architecture: a QRC provides a fast proposal rollout in a reduced-order latent space, and a PINN-based physics-informed corrector (PIC) locally projects the proposal toward lower-residual trajectories. KS is treated as the primary benchmark because it better reflects the chaotic forecasting objective than Burgers.

Our contributions are threefold:
\begin{enumerate}[leftmargin=1.2em]
    \item We formulate a reduced-order \emph{QRC + PIC} pipeline for PDE forecasting in which the QRC predicts proper-orthogonal-decomposition (POD) coefficients and the PIC acts as a local physics-informed residual corrector in reconstructed field space.
    \item We provide an honest two-benchmark evaluation: Burgers as an auxiliary smooth-PDE benchmark and KS as the main chaotic-PDE benchmark.
    \item We show that the physics-informed corrector consistently improves the QRC proposal on the main benchmark, while Burgers highlights an important limitation---in easier reduced-order regimes, simple autoregressive baselines can remain highly competitive.
\end{enumerate}

\section{Related Work}
\subsection{Physics-informed neural networks and their limitations}
PINNs introduced the now-standard recipe of combining supervised or weakly supervised losses with PDE residual penalties computed by automatic differentiation \cite{raissi2019pinns}. The broader scientific machine learning perspective emphasizes that such models can exploit scarce observations and governing equations simultaneously \cite{karniadakis2021piml}. However, benchmark and failure-mode studies have repeatedly shown that naive PINNs become difficult to optimize on convection-dominated, multiscale, or chaotic problems, especially when long horizons or stiff operators are involved \cite{krishnapriyan2021failure,pinnacle2024}. These observations motivate using a PINN-like model as a \emph{local corrector} rather than as the sole global forecaster.

\subsection{Reservoir computing for chaotic and spatiotemporal dynamics}
Echo state networks and related RC models are appealing because only the readout is trained, making them efficient for sequence forecasting \cite{jaeger2001echo}. Pathak \emph{et al.} demonstrated that RC can forecast large spatiotemporally chaotic systems, including KS, using parallelized reduced representations \cite{pathak2018ks}. Physics-informed ESN variants later showed that adding residual-based penalties can improve predictability and hidden-state reconstruction in chaotic systems \cite{doan2020piesn,racca2021apiesn}. This literature is important here because it suggests that hybridizing data-driven rollout models with lightweight physics constraints is a productive design pattern.

\subsection{Quantum reservoir computing}
QRC began as a proposal to use natural or engineered quantum dynamics as nonlinear reservoirs \cite{fujii2017qrc}. Subsequent studies explored alternative physical realizations and mechanisms for QRC, including single-oscillator implementations and dissipation-assisted reservoirs \cite{govia2021oscillator,sannia2024dissipation}. Recent work has made the connection to chaotic forecasting more explicit. Ahmed \emph{et al.} introduced recurrence-free QRC architectures for chaotic dynamics and extreme-event prediction and later analyzed robustness and generalized synchronization properties in quantum reservoirs \cite{ahmed2024extreme,ahmed2025robust}. Other works showed that quantum or hybrid quantum-classical reservoirs can model reduced-order convection and turbulence statistics in fluid systems \cite{pfeffer2022thermal,pfeffer2023rbcrom}. These studies support QRC as a plausible proposal generator for reduced-order PDE states, even if decisive superiority over well-tuned classical baselines remains unproven. Related hybrid quantum--physics-informed models have also been studied for fluid and high-speed-flow settings, although with a different objective from the local-corrector design considered here \cite{sedykh2023hqpinny,leong2025hqpinny}.

\subsection{Physics-informed correction and projection}
A growing line of work studies output-projection or physics-consistent correction mechanisms that take a preliminary forecast and move it toward a physically admissible manifold \cite{valente2025projection}. This is conceptually close to our PIC stage. We do not solve the full PDE from scratch with a PINN; rather, we use the proposal as a strong prior and train a residual corrector whose role is to improve physical consistency without discarding the fast proposal trajectory.

\section{Problem Setup and Benchmarks}
\subsection{Burgers equation}
The auxiliary benchmark is the one-dimensional viscous Burgers equation
\begin{equation}
    u_t + u u_x = \nu u_{xx},
    \label{eq:burgers}
\end{equation}
with periodic boundary conditions and problem-specific initial/forcing choices defined in our data-generation setup. Burgers is a classic PINN benchmark because it combines nonlinear transport and diffusion and can exhibit steep gradients \cite{raissi2019pinns}. In our study, however, it is not the main headline benchmark; instead it functions as a numerically stable, lower-difficulty PDE sanity check.

\subsection{Kuramoto--Sivashinsky equation}
The main benchmark is the KS equation,
\begin{equation}
    u_t + u u_x + u_{xx} + u_{xxxx} = 0,
    \label{eq:ks}
\end{equation}
which is a canonical testbed for spatiotemporal chaos and machine-learning-based forecasting \cite{pathak2018ks,wyder2025ctf}. KS is the more appropriate benchmark when the story is chaotic reduced-order forecasting rather than physics-informed PDE solving in isolation. Accordingly, the paper is organized with KS as the primary benchmark and Burgers as auxiliary.

\subsection{Reduced-order state representation}
For both PDEs we operate in a reduced-order latent space obtained by proper orthogonal decomposition (POD):
\begin{equation}
    u(x,t) \approx \bar{u}(x) + \sum_{i=1}^{r} a_i(t)\,\phi_i(x),
    \label{eq:pod}
\end{equation}
where $\bar{u}$ is the training-set mean field, $\phi_i$ are POD modes learned on the training trajectories only, and $a_i(t)$ are modal coefficients. The QRC, ESN, and classical autoregressive baselines all forecast the same coefficient representation, ensuring a fair state-space comparison across models.

\section{Method}
\subsection{Stage I: Quantum reservoir proposal}
Let $\mathbf{a}_t \in \mathbb{R}^{r}$ denote the reduced-order coefficient vector at time $t$. We construct an input-encoded pure-state QRC with $n_q$ qubits and $V$ virtual nodes. Each normalized input is mapped to phase angles via a fixed random affine map, propagated through a fixed random unitary, and read out through expectation-like features such as $Z_i$, $X_i$, and nearest-neighbor $Z_i Z_{i+1}$. Concatenating the virtual-node features produces a reservoir state $\mathbf{r}_t$, and the next-step proposal is generated by a ridge-regression readout
\begin{equation}
    \hat{\mathbf{a}}_{t+1} = W_{\text{out}}\,\mathbf{r}_t.
    \label{eq:qrc_readout}
\end{equation}
The QRC is trained with teacher forcing, but validation and test metrics are computed under closed-loop rollout.

\subsection{Stage II: PINN-based physics-informed correction}
Given a local proposal window $\hat{\mathbf{a}}_{t:t+H-1}$ and its reconstructed field window $\hat{u}(x,t)$, the corrector predicts a residual-corrected trajectory
\begin{equation}
    \tilde{u}(x,t) = \hat{u}(x,t) + \Delta_\theta(x,t;\hat{u}),
\end{equation}
where $\Delta_\theta$ is a small neural correction. In practice we use a PINN-style corrector and train it on short windows with a composite loss
\begin{equation}
    \mathcal{L} = \lambda_{\text{data}}\,\|\tilde{u}-u\|_2^2
    + \lambda_{\text{prior}}\,\|\tilde{u}-\hat{u}\|_2^2
    + \lambda_{\text{phys}}\,\|\mathcal{F}(\tilde{u})\|_2^2,
    \label{eq:picloss}
\end{equation}
where $u$ is the true field on training/validation windows only, $\hat{u}$ is the proposal, and $\mathcal{F}$ is the PDE residual operator corresponding to Eq.~\eqref{eq:burgers} or Eq.~\eqref{eq:ks}. At test time the corrector sees only the proposal windows. In other words, the correction is autonomous and does not rely on hidden test-time truth anchoring.

\subsection{Why proposal--correction rather than PINN-only?}
Global PINN-only training can be brittle for chaotic long-horizon PDE forecasting, while efficient proposal models such as RC are often more stable in practice \cite{pathak2018ks,doan2020piesn,krishnapriyan2021failure,pinnacle2024}. We therefore adopt a proposal--correction split so the PINN stage focuses on local physics-consistency refinement rather than full trajectory generation.

\section{Experimental Protocol}
\subsection{Splits, seeds, and model selection}
All reported experiments follow the same protocol:
\begin{itemize}[leftmargin=1.2em]
    \item trajectory-level train/validation/test split;
    \item training-set-only POD basis and normalization statistics;
    \item five random seeds for stochastic models;
    \item hyperparameter selection on validation rollout only;
    \item final reporting on held-out test trajectories only.
\end{itemize}
This design follows a strict evaluation discipline: fixed rollout conditions, validation-only tuning, and comparison of proposal-only against proposal+correction models under the same reduced-order representation.

\subsection{Baselines}
We compare against a compact but informative baseline set:
\begin{itemize}[leftmargin=1.2em]
    \item Persistence,
    \item Ridge autoregression (Ridge-AR),
    \item MLP autoregression (MLP-AR),
    \item PI-MLP (a PINN-like autoregressive baseline),
    \item ESN,
    \item ESN+PIC,
    \item QRC,
    \item QRC+PIC.
\end{itemize}
The PI-MLP baseline is conceptually related to physics-constrained autoregressive models for PDE dynamics forecasting \cite{geneva2020pde}.
The resulting comparison isolates three questions: (i) whether QRC is competitive as a proposal generator, (ii) whether a physics-informed corrector improves proposal quality, and (iii) whether any quantum benefit survives comparison with strong classical reduced-order baselines.

\subsection{Metrics}
For each benchmark we report mean $\pm$ standard deviation across seeds for field-level RMSE, NRMSE, and PDE residual. Where available, we also report valid prediction time (VPT), medium-horizon and long-horizon RMSE, and a spectral mismatch score. We emphasize two complementary axes: predictive accuracy and physical consistency.

\begin{figure}[t]
\centering
\begin{subfigure}[t]{0.5\textwidth}
\centering
\includegraphics[width=\linewidth]{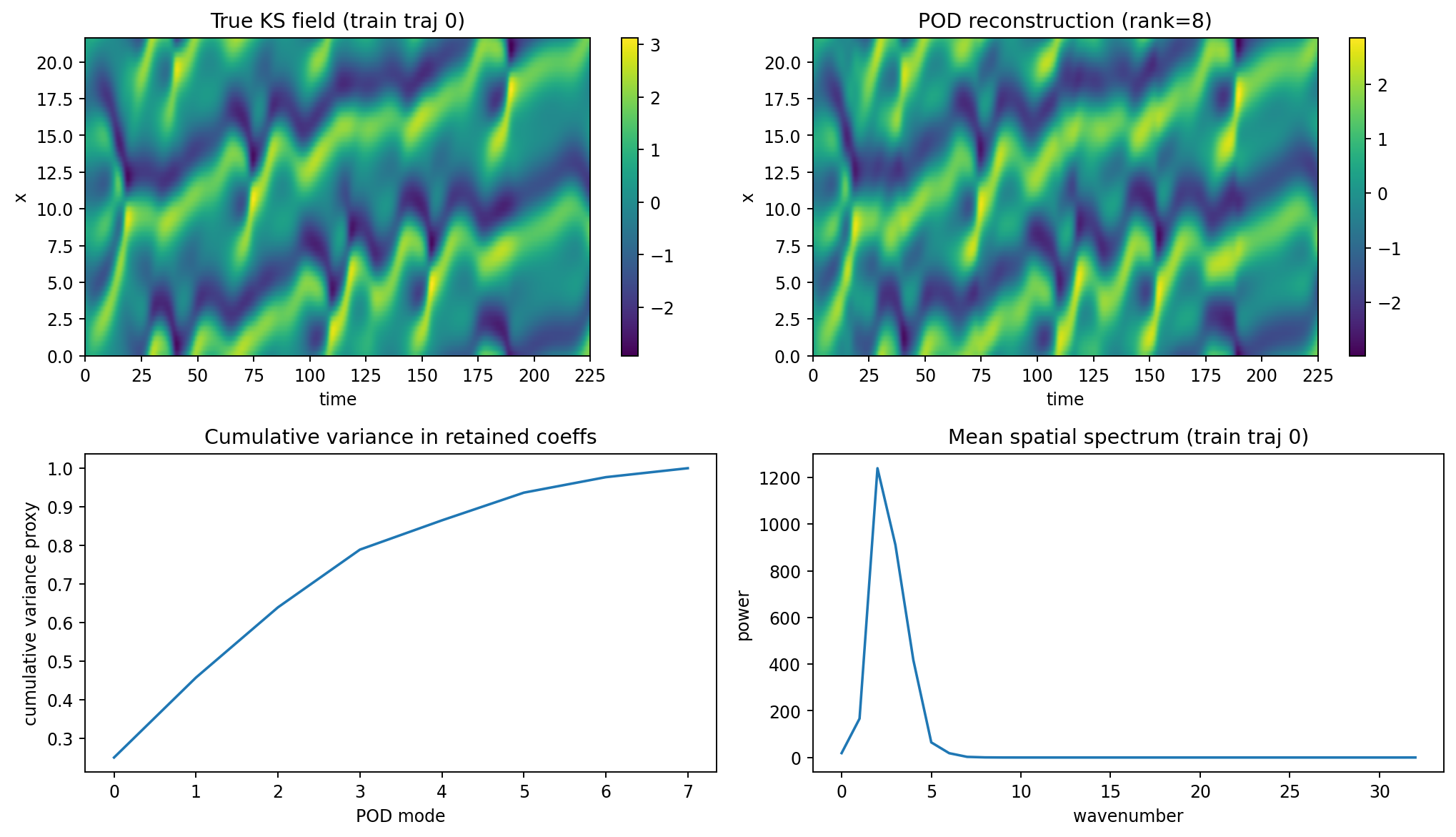}
\caption{Kuramoto--Sivashinsky (primary).}
\end{subfigure}\hfill
\begin{subfigure}[t]{0.5\textwidth}
\centering
\includegraphics[width=\linewidth]{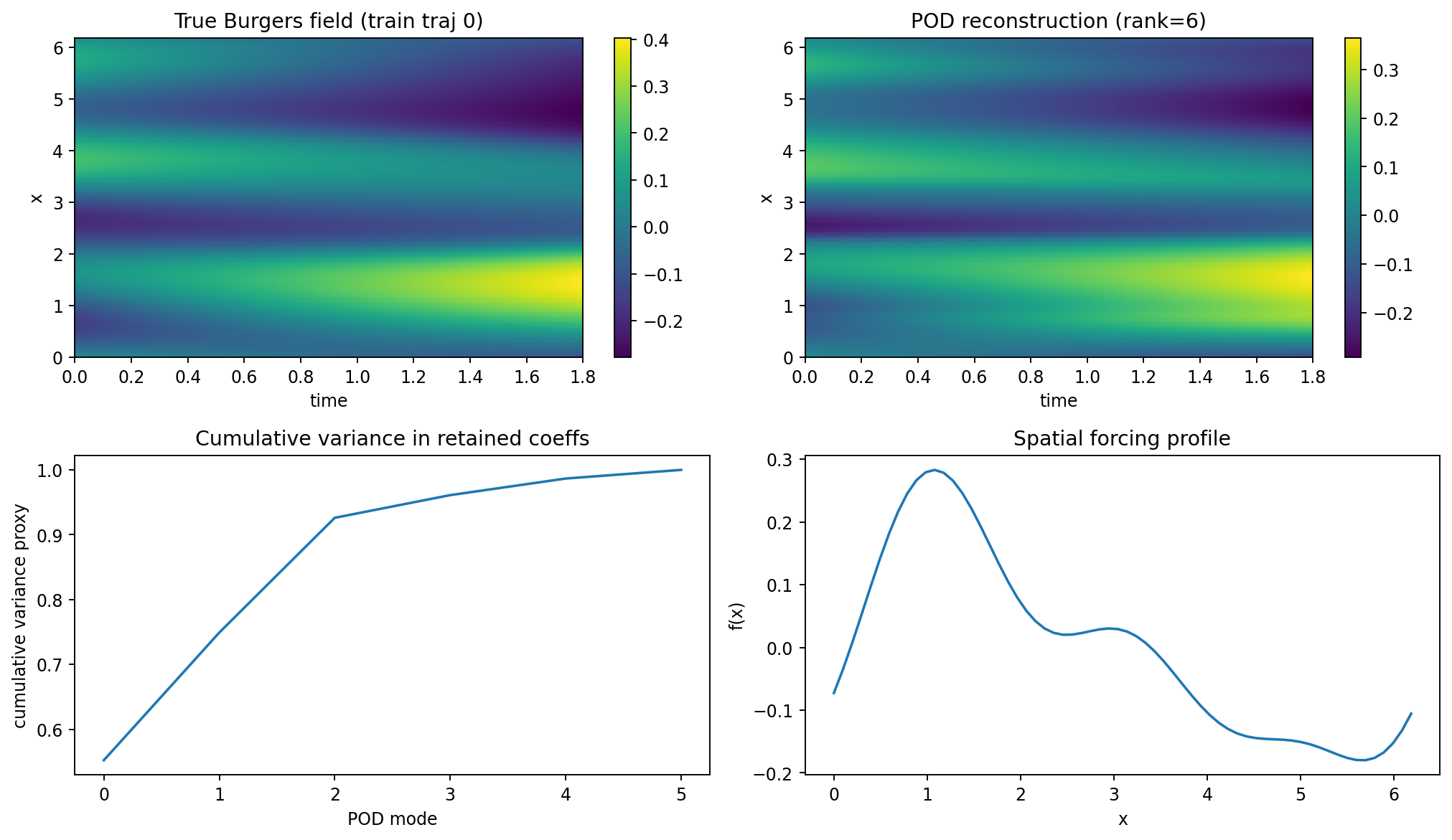}
\caption{Burgers (auxiliary).}
\end{subfigure}
\caption{Dataset overview figures}
\label{fig:dataset-overview}
\end{figure}

Figure~\ref{fig:dataset-overview} shows dataset-level diagnostics for the KS and Burgers benchmarks.

\subsection{QRC+PIC Pipeline}

\begin{figure}[H]
\centering
\resizebox{\textwidth}{!}{
\begin{tikzpicture}[
box/.style={draw, rectangle, rounded corners, align=center, minimum width=2.6cm, minimum height=0.9cm},
arrow/.style={->, thick}
]

\tikzstyle{data}=[box, fill=blue!20]
\tikzstyle{model}=[box, fill=green!25]
\tikzstyle{physics}=[box, fill=orange!30]
\tikzstyle{output}=[box, fill=purple!25]

\node[data] (data) at (0,0) {PDE Data};
\node[model] (train) at (3,0) {Train QRC};
\node[model] (qrc) at (6,0) {QRC Model};

\node[model] (rollout) at (3,-2) {Closed-Loop\\Rollout};
\node[physics] (pinn) at (6,-2) {PIC\\Correction};
\node[output] (output) at (9,-2) {Final\\Solution};

\draw[arrow] (data) -- (train);
\draw[arrow] (train) -- (qrc);

\draw[arrow] (qrc) -- (rollout);
\draw[arrow] (rollout) -- (pinn);
\draw[arrow] (pinn) -- (output);

\end{tikzpicture}
}
\caption{Workflow of the proposed QRC+PIC framework. A quantum reservoir generates rollout predictions that are then refined by a physics-informed corrector.}
\label{fig:qrc-pic-pipeline}
\end{figure}
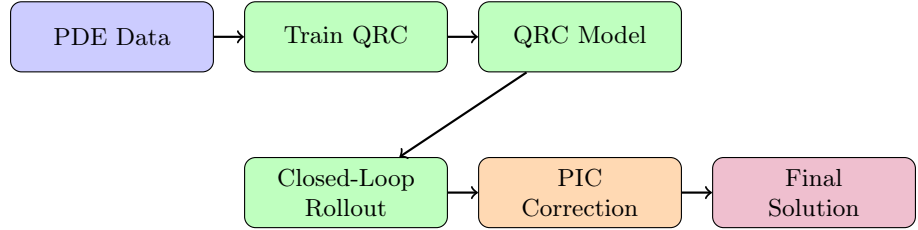

\section{Results}
\subsection{Primary benchmark: Kuramoto--Sivashinsky}
Table~\ref{tab:ks-main} summarizes the main KS results. Two observations matter most. First, the physics-informed correction consistently improves the QRC proposal: QRC+PIC improves QRC in RMSE, NRMSE, and physics residual. Second, the margins against strong simple baselines are small.

\begin{table}[H]
\caption{Main KS results (five seeds, test trajectories only)}
\label{tab:ks-main}
\centering
\small
\begin{tabular}{lccc}
\toprule
Model & RMSE $\downarrow$ & NRMSE $\downarrow$ & Physics residual $\downarrow$ \\
\midrule
Ridge-AR & 1.1777 $\pm$ 0.0000 & 1.0100 $\pm$ 0.0000 & 0.6331 $\pm$ 0.0000 \\
QRC+PIC & 1.1897 $\pm$ 0.0027 & 1.0204 $\pm$ 0.0024 & 0.7035 $\pm$ 0.0498 \\
QRC & 1.1971 $\pm$ 0.0044 & 1.0268 $\pm$ 0.0039 & 0.7830 $\pm$ 0.0367 \\
Persistence & 1.2606 $\pm$ 0.0000 & 1.0807 $\pm$ 0.0000 & 0.0715 $\pm$ 0.0000 \\
ESN+PIC & 1.5009 $\pm$ 0.0604 & 1.2871 $\pm$ 0.0516 & 2.8606 $\pm$ 0.4891 \\
PI-MLP & 1.6390 $\pm$ 0.1326 & 1.4057 $\pm$ 0.1143 & 2.3110 $\pm$ 0.5685 \\
ESN & 1.7070 $\pm$ 0.0484 & 1.4639 $\pm$ 0.0411 & 4.9063 $\pm$ 1.0826 \\
MLP-AR & 1.7192 $\pm$ 0.0671 & 1.4748 $\pm$ 0.0576 & 4.2180 $\pm$ 1.3288 \\
\bottomrule
\end{tabular}
\end{table}

In our view, the correct interpretation is not that the quantum model has established superiority, but that the hybrid proposal--correction design is functioning as intended on a genuinely chaotic PDE. QRC+PIC improves QRC, and it substantially outperforms the ESN-family baselines in this configuration. The fact that Ridge-AR remains slightly stronger is scientifically important rather than embarrassing: it shows that reduced-order chaotic forecasting can remain surprisingly linear in some regimes and that hybrid quantum models must be benchmarked conservatively.

\begin{figure}[t]
\centering
\begin{subfigure}[t]{0.8\textwidth}
\centering
\includegraphics[width=\linewidth]{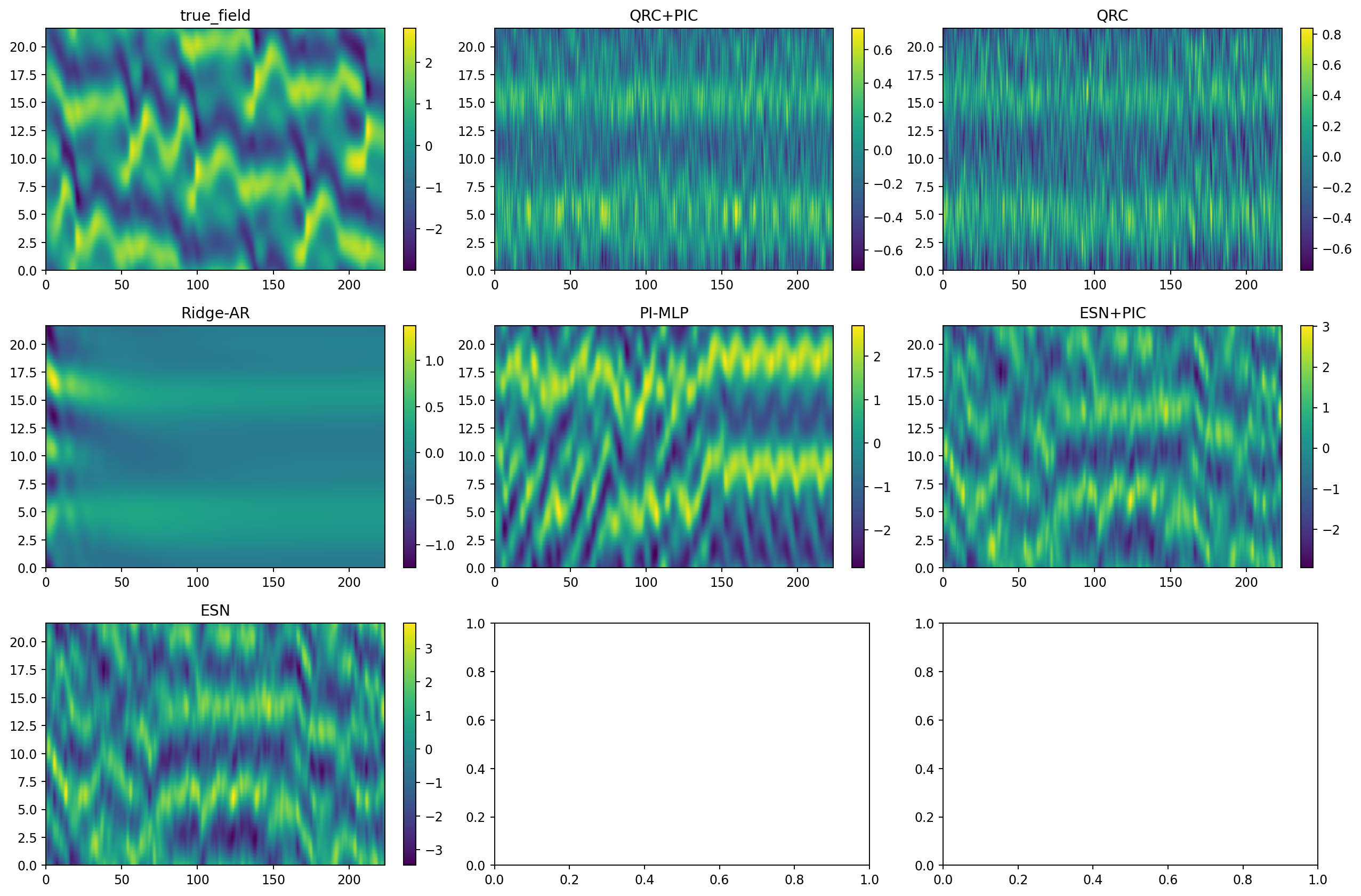}
\caption{Space-time rollout comparison.}
\end{subfigure}

\begin{subfigure}[t]{0.8\textwidth}
\centering
\includegraphics[width=\linewidth]{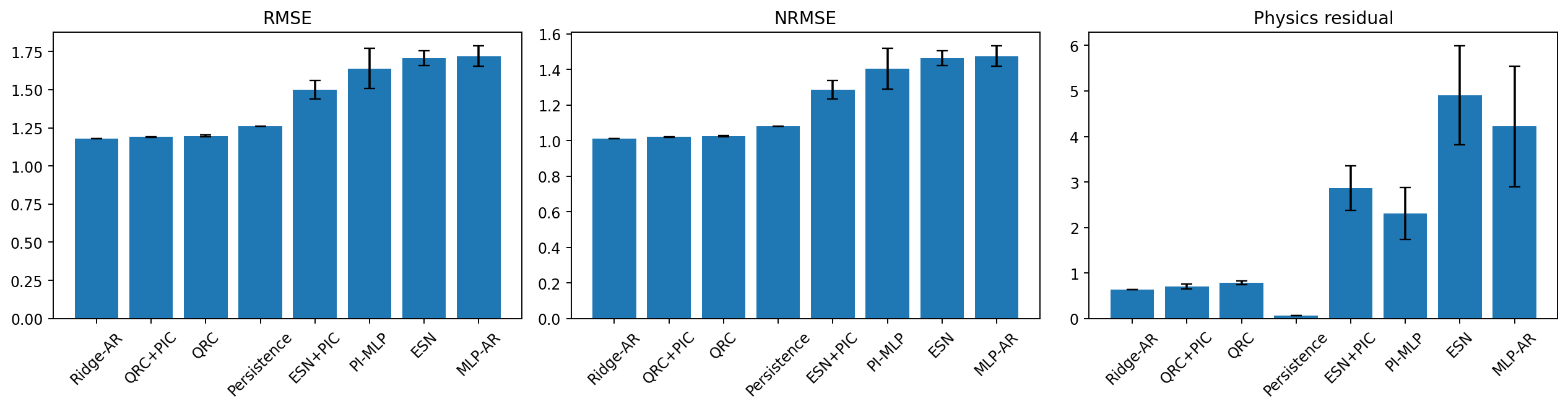}
\caption{Aggregate test metrics with seed-wise error bars.}
\end{subfigure}
\caption{Primary KS figures.}
\label{fig:ks-main-figures}
\end{figure}

Figure~\ref{fig:ks-main-figures} shows representative KS rollout comparisons and aggregate metric bars.

\subsection{Auxiliary benchmark: Burgers}
Table~\ref{tab:burgers-main} summarizes the auxiliary Burgers results. In this smoother reduced-order regime, persistence and simple autoregressive baselines remain strong. QRC+PIC improves over QRC alone, but Burgers does not provide the clearest setting for arguing that the reservoir is essential. We therefore use Burgers as an auxiliary sanity-check benchmark rather than the main result.

\begin{table}[t]
\caption{Auxiliary Burgers results (five seeds, test trajectories only).}
\label{tab:burgers-main}
\centering
\small
\begin{tabular}{lccc}
\toprule
Model & RMSE $\downarrow$ & NRMSE $\downarrow$ & Physics residual $\downarrow$ \\
\midrule
Ridge-AR & 0.0644 $\pm$ 0.0000 & 0.2025 $\pm$ 0.0000 & 0.0298 $\pm$ 0.0000 \\
Persistence & 0.1887 $\pm$ 0.0000 & 0.5836 $\pm$ 0.0000 & 0.0782 $\pm$ 0.0000 \\
QRC+PIC & 0.3458 $\pm$ 0.0010 & 1.0647 $\pm$ 0.0031 & 15.0128 $\pm$ 3.4132 \\
MLP-AR & 0.3573 $\pm$ 0.0030 & 1.1164 $\pm$ 0.0089 & 1.2029 $\pm$ 0.0411 \\
QRC & 0.3578 $\pm$ 0.0040 & 1.1025 $\pm$ 0.0130 & 177.5869 $\pm$ 35.5570 \\
PI-MLP & 0.3861 $\pm$ 0.0436 & 1.2114 $\pm$ 0.1429 & 1.3334 $\pm$ 0.1445 \\
ESN+PIC & 0.4397 $\pm$ 0.1148 & 1.3754 $\pm$ 0.3572 & 12.7754 $\pm$ 4.9017 \\
ESN & 0.5882 $\pm$ 0.1568 & 1.8436 $\pm$ 0.4906 & 22.0871 $\pm$ 7.7836 \\
\bottomrule
\end{tabular}
\end{table}

This benchmark therefore plays an auxiliary role in the paper. It verifies that the Burgers data generator, reduced-order representation, and physics-informed correction machinery are operational, but it also highlights a limitation: in benign reduced-order regimes, a hybrid quantum proposal is not automatically competitive with simple local baselines.

\begin{figure}[t]
\centering
\includegraphics[width=0.5\linewidth]{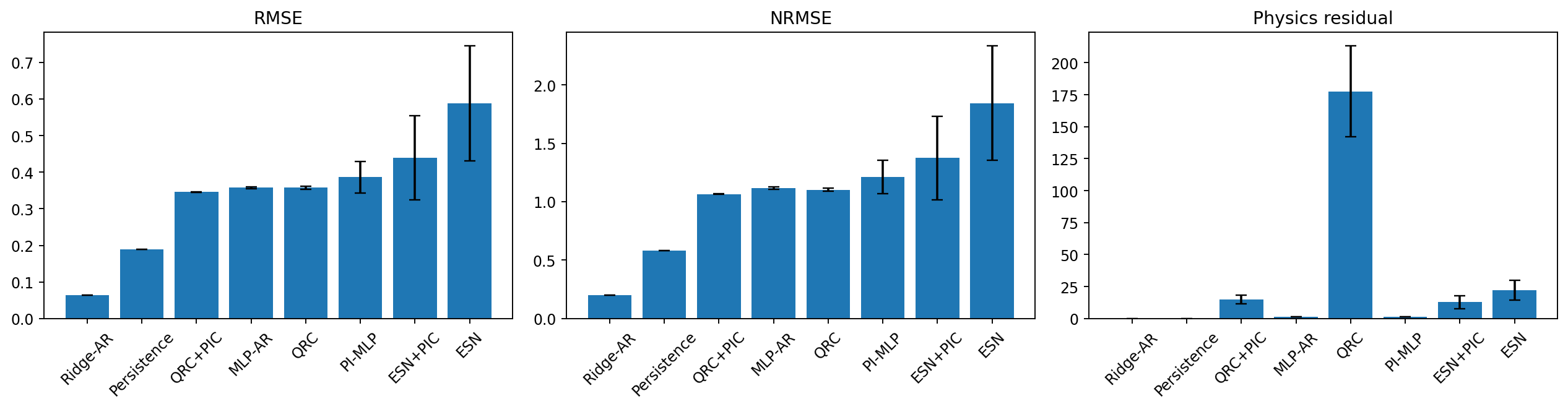}
\caption{Aggregate test metrics with seed-wise error bars for Burgers}
\label{fig:burgers-main-figures}
\end{figure}

Figure~\ref{fig:burgers-main-figures} shows representative Burgers rollout comparisons and aggregate metric bars.

\subsection{Ablation and sensitivity analysis}
\begin{figure}[t]
\centering
\includegraphics[width=\textwidth]{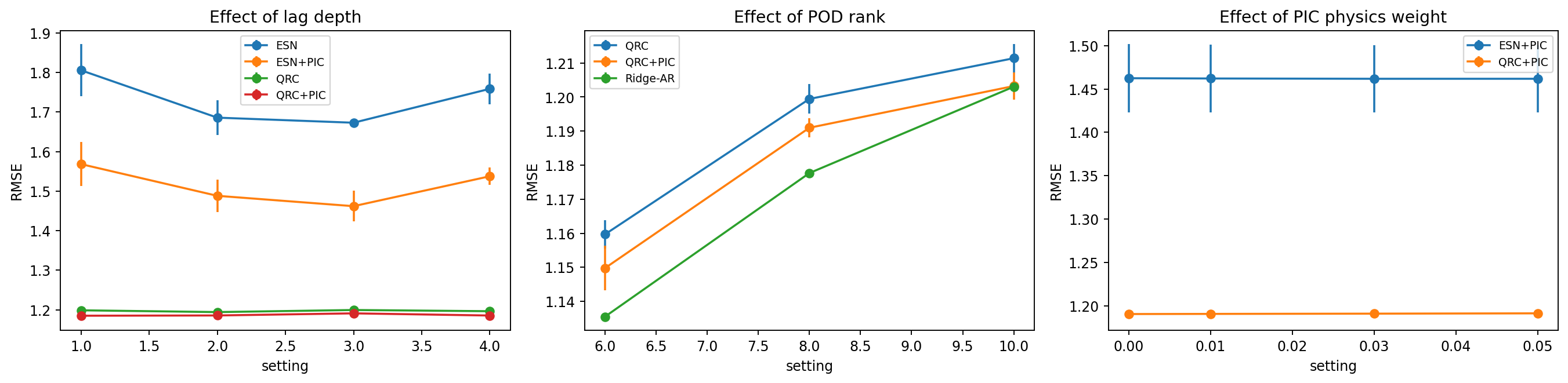}
\caption{KS ablation overview, including lag depth, POD rank, horizon, and physics-loss sensitivity.}
\label{fig:ks-ablation}
\end{figure}

The most practically important ablations are (i) QRC versus QRC+PIC, (ii) ESN versus ESN+PIC, (iii) lag depth, (iv) POD rank, (v) rollout horizon, and (vi) PIC physics-loss weight. Figure~\ref{fig:ks-ablation} summarizes these sweeps. The trends confirm that correction quality is sensitive to configuration and loss balancing, which aligns with the failure-mode literature on PINNs \cite{krishnapriyan2021failure,pinnacle2024}.

\section{Discussion}
Several conclusions emerge.

\paragraph{1. The proposal--correction idea is viable but benchmark-dependent.}
The main KS result supports the central architectural claim: a QRC proposal can be improved by a local physics-informed corrector, and the defensible novelty is the proposal--projection hybrid rather than a ``quantum PINN'' monolith.

\paragraph{2. Burgers should remain auxiliary.}
Burgers is best treated as an auxiliary benchmark rather than the flagship result. It remains useful for debugging, visualization, and sanity checks, but the main chaotic forecasting story belongs to KS.

\paragraph{3. Strong classical baselines must remain in the paper.}
Ridge-AR is a particularly important baseline because it demonstrates that low-dimensional reduced-order dynamics can remain surprisingly linear. Omitting it would weaken the credibility of the study. The right narrative is not that QRC+PIC dominates all baselines; it is that QRC+PIC is a competitive hybrid with a favorable accuracy--physics tradeoff and a clear improvement over QRC alone on the main benchmark.

\paragraph{4. The PIC is better viewed as a local physics-informed corrector than as a full PINN solver.}
The architecture is best described as \emph{QRC + PINN-based physics-informed correction} or \emph{QRC + PIC}, rather than as a PINN that directly solves the full PDE from scratch. The latter would overstate the role of the second stage.

\section{Limitations and Threats to Validity}
The study has several limitations. First, the best overall KS result remains close to a simple linear baseline. Second, the auxiliary Burgers benchmark does not showcase a strong quantum advantage and instead confirms that some smooth reduced-order regimes are well served by persistence or linear AR models. Third, the QRC used here is a pure-state simulated reservoir rather than a hardware execution or noisy quantum device.

\section{Conclusion}
We presented a reduced-order PDE forecasting pipeline in which a pure-state quantum reservoir computer generates a closed-loop proposal rollout and a PINN-based physics-informed corrector refines the proposal locally in field space. The main result is on the Kuramoto--Sivashinsky equation, where the corrector consistently improves the QRC proposal and yields a competitive accuracy--physics tradeoff under multi-seed, validation-only evaluation. The Burgers equation serves as an auxiliary benchmark that verifies the robustness of the pipeline while showing that easy reduced-order regimes can still favor trivial local baselines. Overall, the evidence supports a cautious but positive conclusion: \emph{QRC proposals with local physics-informed correction are a viable reduced-order SciML architecture for nonlinear PDE forecasting, especially when the goal is not headline quantum advantage but physically consistent competitive forecasting under tight computational budgets.}

\end{document}